 \documentstyle[twoside,fleqn,    npbmod,epsfig]{article}

\def\gsim{\lower0.5ex\hbox{$\stackrel{>}{\sim}$}}
\def\lsim{\lower0.5ex\hbox{$\stackrel{<}{\sim}$}}

\def\xfigti
{
\begin{figure}[t]
 \epsfxsize=0.75\hsize
 \centerline{\epsfbox{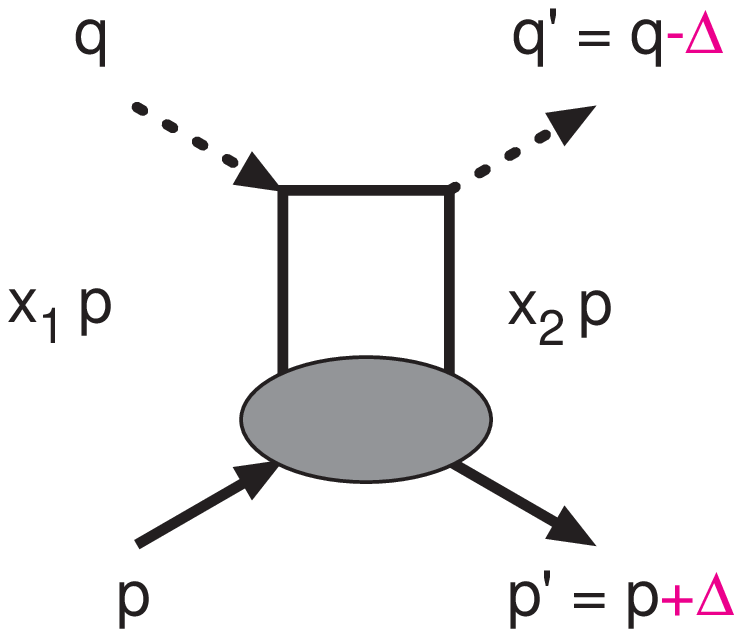}}
 \vspace{-0.50in}
 \caption{The kinematics of the DVCS process.}
 \label{fig:}
 \vspace{-0.25in}
\end{figure}
}
\def\xfigregge
{
\begin{figure}[t]
 \epsfxsize=0.75\hsize
 \centerline{\epsfbox{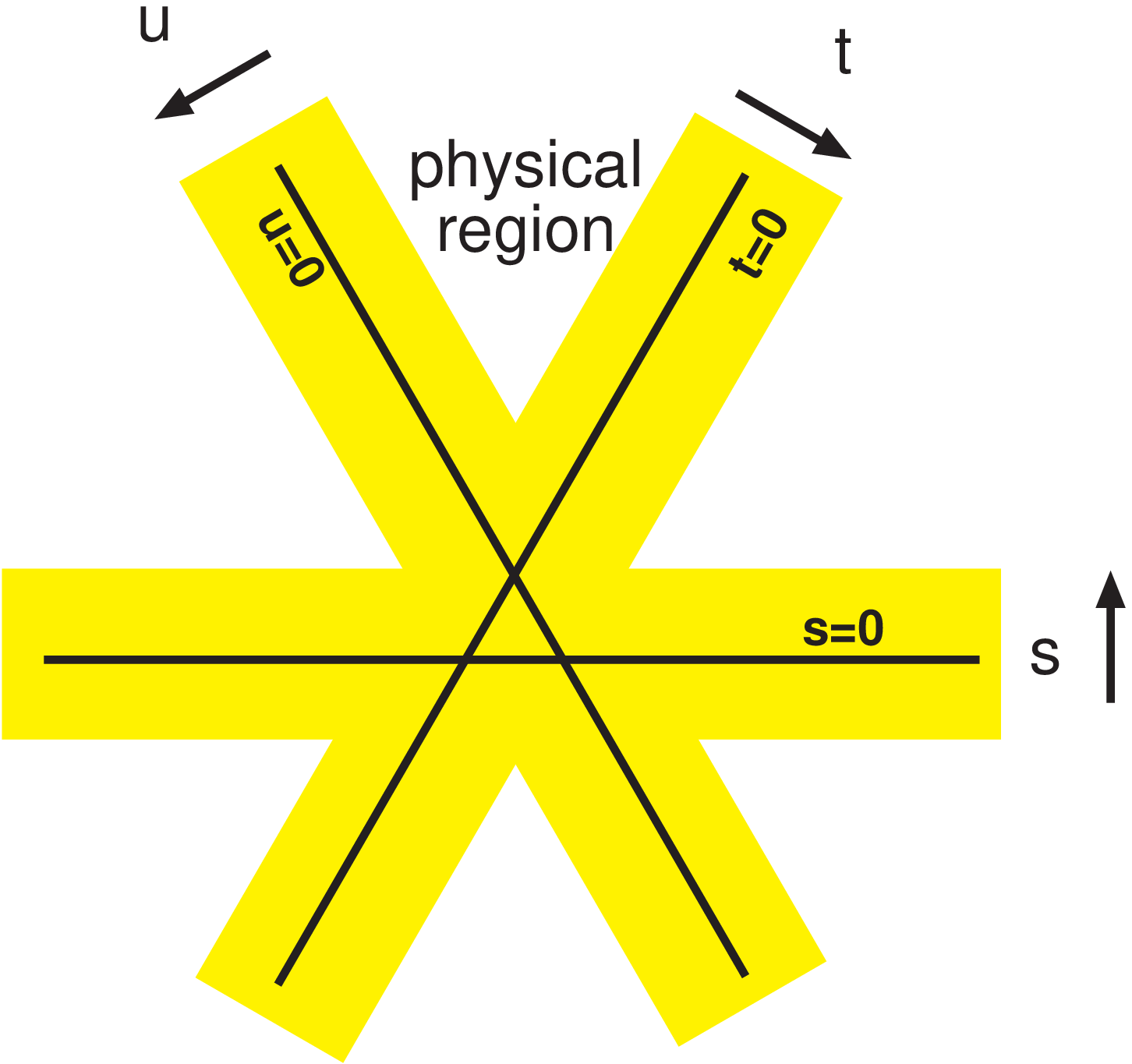}}
 \vspace{-0.25in}
 \caption{The Mandelstam diagram showing the 
{\it strip regions} (shaded) where the amplitude can be 
accurately described by a simple pole structure.}
 \label{fig:}
 \vspace{-0.25in}
\end{figure}
}
\def\xfigv
{
\begin{figure}[t]
 \epsfxsize=0.95\hsize
 \centerline{\epsfbox{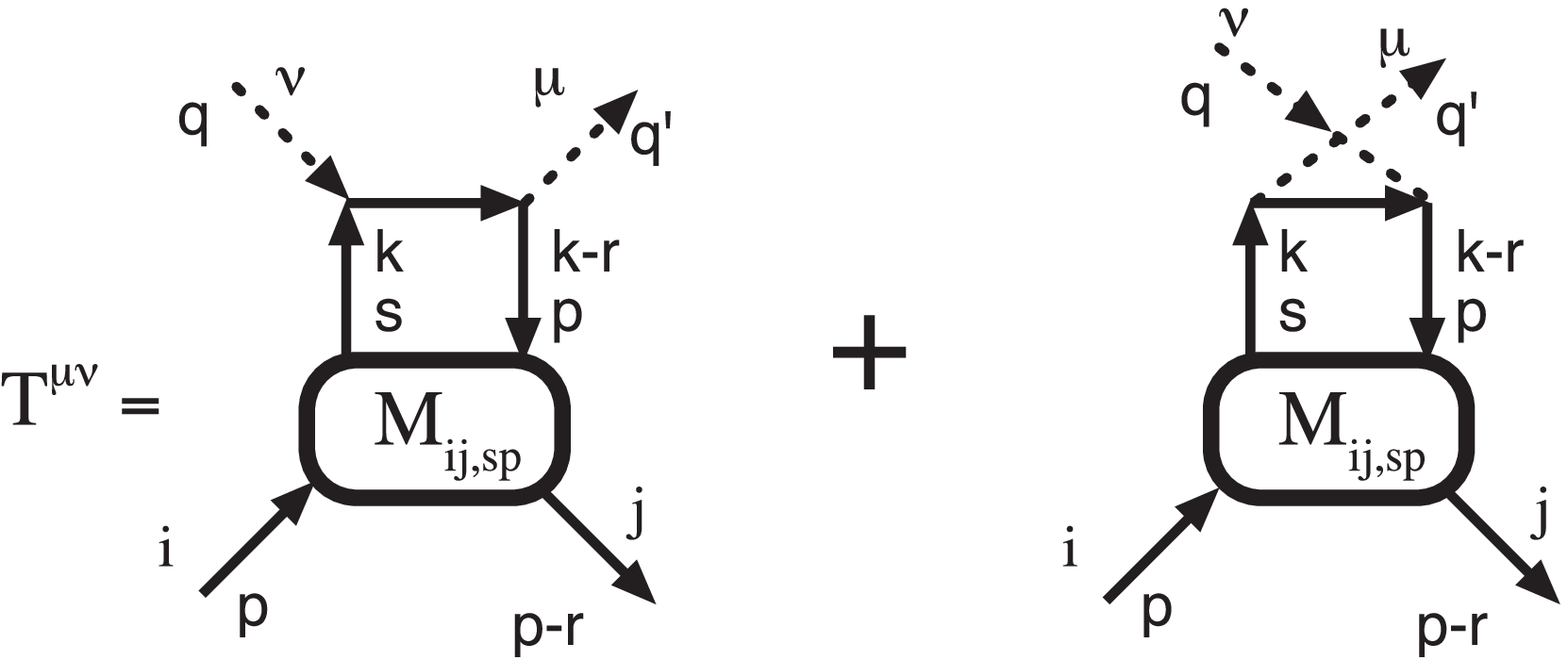}}
 \vspace{-0.25in}
 \caption{Total amplitude for the DVCS process.}
 \label{fig:}
 \vspace{-0.25in}
\end{figure}
}

\title{
Asymptotic properties of DVCS\thanks{Presented by F. Olness.  
 To be published in the proceedings of 7th International Workshop on
Deep Inelastic Scattering and QCD (DIS 99), Zeuthen, Germany, 19-23
Apr 1999. 
 This work is supported in part by
 Grant INTAS-RFBR-95-311,
 Grant RFBR-98-02-17629,
 the US Department of Energy, and the
Lightner-Sams Foundation. 
 We thank V.Teplitz for helpful discussion.
}
}

\author{
B.I.~Ermolaev\address{A.F.~Ioffe Physico-Technical Institute, St.~Petersburg, 194021, Russia}
F.~Olness\address{Southern Methodist University, Dallas, TX 75275, USA}
and
A.G.~Shuvaev\address{St.~Petersburg Nuclear Physics Institute, Gatchina, St.~Petersburg, 188350, Russia }
}

\begin{document}

\begin{abstract}

 We compute  the  deeply virtual Compton scattering  (DVCS) amplitude for
forward and backward scattering in the asymptotic limit.
 We make use of the Regge calculus to
resum  important logarithmic contributions that are
beyond those included by the DGLAP evolution.
 We find a power-like behavior for the forward DVCS amplitude.
\end{abstract}

\def\pretitle{\hbox to \hsize{\hfill {\normalsize hep-ph/9906294}}}

\maketitle
\section{Deeply Virtual Compton Scattering}

The deeply inelastic scattering (DIS) process, $\gamma^* P \to \gamma^* P$ 
has provided a wealth of information about the structure of the 
proton, or more specifically, the hadronic matrix element: $\langle P | ... | P \rangle$. 
 Recently, it was realized that one can actually investigate the non-diagonal 
 hadronic matrix element $\langle P' | ... | P \rangle$ via the 
Deeply Virtual Compton Scattering (DVCS) process $\gamma^* P \to \gamma   P'$. 
 This DVCS process differs from the DIS process in that the outgoing photon is on-shell,
and hence the final hadron $P'$  must absorb a small amount of momentum $\Delta$ to satisfy 
the kinematics.\cite{Ji,Rad,BV,DVCS,DIS99}
 The net result of the DVCS analysis is to obtain a generalized parton distribution function (PDF) 
which depends on two scaling variables, $\{x_1,x_2 \}$, {\it cf.}, Fig.1. 
Various sum rules can be derived to relate these DVCS 
distributions $f(x_1,x_2)$ to the usual PDF's, $f(x)$.\cite{DVCS}
 \xfigti
 %

\section{Regge Analysis}

We want to analyze this new DVCS sub-field 
using some  older techniques, namely Regge theory, 
to obtain analytic expressions for the scattering amplitudes. 
 The key idea that we will use from  Regge analysis is that if we confine 
our kinematics to the  {\it strip regions} of the Mandelstam diagram (Fig.2), then 
the amplitude can accurately be described by a simple pole structure. 
If we stray beyond these strip regions into the rest of the physical region, then
we are too far from low-energy crossed-channel poles to be
dominated by them.
 \xfigregge
Therefore, we will limit our analysis to  one of two regions:
\\ \null \quad 1) the Forward  Region: $s \sim - u \gg -t$, and 
\\ \null \quad 2) the Backward Region: $s \sim - t \gg -u$,
\\
with, of course $s \gg Q^2$.

\section{Infrared Evolution Equation: (IREE)}

Since the higher order QCD corrections to the DVCS process (Fig.1) 
do not violate the Born spin structure, the general DVCS amplitude is: 
$M_F  = M_F^{Born} \ \phi[z_s, z_Q, z_t]$, with:
 \begin{eqnarray*}
 \begin{array}{lcllcl}
z_s &=&  \ln(s/\mu^2)   &
z_t &=&  \ln(-t/\mu^2)  \\
z_Q &=&  \ln(Q^2/\mu^2) &
z_u &=&  \ln(-u/\mu^2)   \\
 \end{array} 
 \end{eqnarray*}
We differentiate w.r.t. $\mu$ to obtain the Infrared Evolution Equation (IREE):
 \begin{eqnarray*}
-\mu^2 \
  \frac{\partial M_F}{\partial \mu^2}
&=&
  \frac{\partial M_F}{\partial z_s}
+ \frac{\partial M_F}{\partial z_Q}
+ \frac{\partial M_F}{\partial z_t}
 \end{eqnarray*}
Analysis of $M_F$  yields the solution:
 \begin{eqnarray*}
M_F &=& \phi (z_s - z_t, z_Q - z_t) \ e^{-\, \frac{\alpha_s C_F}{4\pi}  \, z_t^2}
\label{eq:eleven}
 \end{eqnarray*}

\section{Forward Region: $s  > Q^2 > -t \gg \mu^2$}

 We decompose $M_F$ into spin-dependent and spin-independent terms, 
$M_F = N_F + U_F.$
 We first obtain the spin-dependent asymptotic behavior in the 
Regge limit (large s) for the forward region ($t\simeq 0$):
 \begin{eqnarray*}
N_F &\simeq&
\left( \frac{s}{Q^2} \right)^a \
\left( \frac{Q^2}{\mu^2} \right)^{a/2} \
e^{-\frac{\alpha_s}{4\pi}\ln^2(-t/\mu^2)}
\label{eq:twentyseven}
 \end{eqnarray*}
with
$a \simeq 3.5 [2\alpha_s N/(4\pi]^{1/2}$

In a similar manner, for the spin-independent part, we obtain:
 \begin{eqnarray*}
U_F &\simeq& 
\left( \frac{s}{Q^2} \right)^{1 + \Delta_P} \
\left( \frac{Q^2}{\mu_2} \right)^{\gamma_P}
\ e^{-\frac{\alpha_sN}{4\pi}\ln^2(-t/\mu^2)}
 \end{eqnarray*}
where $\Delta_P$ is the Pomeron intercept, 
and $\gamma_P$ is the Pomeron anomalous dimension.

Consequently, we find the asymptotic forward scattering  amplitude
takes the form of the Compton scattering amplitude times a Sudakov
exponential factor.

\section{Backward Region: $s  > Q^2 > -u \gg \mu^2$}

For the backward DVCS process ($u\simeq 0$), the analysis is simpler because  
there is no difference between asymptotic behavior of the
polarized and unpolarized  amplitude $M_B$.
 We obtain at the final expression:
$$
M_B = M_B^{Born} \ e^{-\frac{\alpha_s}{4\pi} \, C_F  \ln^2(s/Q^2)} 
$$
 The DVCS amplitude in the backward scattering region is  purely
of the Sudakov type with an
exponential factor.

\section{Relation to DGLAP Analysis}

Here we highlight some differences of this work in comparison to the 
DGLAP analysis. 
For the calculation of the spin-dependent amplitude $N_F$ in the
$-t \sim 0$ limit, we define $F$ via  a Mellin transform: 
 \begin{eqnarray*}
N_F(z_s, z_Q) &=& \int\limits_{-\imath\infty}^{\imath\infty} \
\frac{d\omega}{2\pi\imath} \ e^{z_s\omega} \ \xi(\omega)  \
F(\omega, z_Q)
 \end{eqnarray*}
 The IREE for the spin-dependent part takes the following form:
 \begin{eqnarray*}
\left(\frac{\partial}{\partial z_Q}  + \omega\right) F(\omega, z_Q)
&=&
\frac{1}{8\pi^2} F(\omega, z_Q) f_0(\omega)
 \end{eqnarray*}
where
$f_0$ is the Mellin amplitude for quark-quark elastic scattering  with
all quarks on-shell.
Putting all the pieces together, we obtain for $N_F$:
 \begin{eqnarray*}
 \frac{e^2}{g^2 C_F} 
\int\limits_{-\imath \infty}^{\imath \infty} \frac{d\omega}{2\pi\imath}
\left( \frac{s}{Q^2} \right)^{\omega} 
\frac{e^{-\, \frac{\alpha_s C_F}{4\pi}\, z_t^2}}{\omega - f_0/8\pi^2} 
\left( \frac{Q^2}{\mu^2} \right)^{\frac{f_0}{8\pi^2}} 
 \end{eqnarray*}
 Note that if we expand $f_0$  keeping the leading powers of $1/\omega$,
we arrive at
the standard  leading-order DGLAP leading order expression for the DVCS. 
 The DGLAP incorporates  both singular and  non-singular contributions
from a finite number of terms. 
 Here, we instead resum only the singular contributions
from an infinite number of terms to double-log accuracy. 
While we neglect the  nonsingular terms, 
this does not affect the asymptotic behavior.\cite{BFKL}
 \xfigv
 %

\section{Non-forward structure functions}

Now that we have expressions for the asymptotic forward and 
backward amplitudes, $M_F$ and $M_B$, we can relate these quantities to the
total amplitude which, for $t\simeq 0$ is given by:
 \begin{eqnarray*}
T^{\mu \nu} &=& \frac{g^{\mu \nu}_\perp}{2p \cdot q^\prime} \ T_1 \ - \
\frac{i\varepsilon^{\mu \nu \rho \eta}}{p \cdot q^\prime}  
\frac{p_{\rho} \, q'_{\eta}}{p \cdot q^\prime}   \ T_2    ,
 \end{eqnarray*}
{\it cf.}, Fig.3.
  Using the Sudakov representation for the kinematic variables, we 
can easily evaluate $T_1$ and $T_2$ in the asymptotic limit, and then
invert these equations to obtain analytic expressions for 
the  non-forward structure functions
${\cal F}^{q,\bar q}_\zeta(x)$ and $G^{q,\bar q}_\zeta(x)$.


We find the following non-forward structure functions:\cite{eos}
\begin{eqnarray*}
\begin{array}{l}
{\cal F}^q_\zeta(\beta) =  \frac{C_F}{8 \pi^2}  
\frac 1{1-\beta}  \int\limits_{-\imath \infty}^{+\imath \infty} 
\frac{d\omega}{2\pi\imath}  R_S(\omega) \frac 1\omega 
\times \hfill
 \\[15pt]
\left[\theta\left(\beta-\frac{\zeta}{2}\right) \left(\frac{1-\beta}\beta \right)^\omega
+ \theta\left(\frac{\zeta}{2}-\beta\right)  \left(\frac{1-\beta}{\zeta -\beta}
\right)^\omega \right]
\end{array} 
\label{eq:fquark}
\end{eqnarray*}
\begin{eqnarray*}
\begin{array}{l}
{\cal F}^{\bar q}_\zeta(\beta)  =  \frac{C_F}{8 \pi^2} 
\frac{1}{1+\beta-\zeta} 
\int\limits_{-\imath \infty}^{+\imath \infty} 
\frac{d\omega}{2\pi\imath}  R_U(\omega)  \frac{1}{\omega}
\times
 \\[15pt]
\left[\theta\left(\beta-\frac{\zeta}{2}\right)  \left(\frac{1+\beta-\zeta}\beta
\right)^\omega
+ \theta\left(\frac{\zeta}{2}-\beta\right)  \left(\frac{1+\beta-\zeta}
{\zeta -\beta}\right)^\omega \right]
\end{array} 
\end{eqnarray*}
 The corresponding $G^{q, \bar q}_\zeta(\beta)$  structure functions are expressed  
via analogous formulas. 
 Here, $R_{S,U}$ are the standard $s$- and $u$-channel Regge amplitudes. 
 The Regge limit implies the argument of the Mellin transform is
be asymptotically large, which corresponds to the the region where
$\zeta \ll 1$ and $\beta \sim \zeta$.
 Note that the pair of $\theta$-functions yields the full support in the variable $\beta$.

\section{Conclusions}
 
 Deeply Virtual Compton Scattering
(DVCS) provides a means to expand our knowledge of the hadron structure and 
extract information on new quantities such as the non-diagonal hadronic matrix element:
 $\langle P' | ... | P \rangle $. 
 Via the IREE, we have resummed singular terms in $1/\omega$ to double
logarithmic accuracy. 
While we have neglected the non-singular terms ({\it eg.}, as included in the 
DGLAP approach), these do not contribute in the asymptotic region. 
 This allows us to obtain expressions for the structure functions 
in the forward and backward regions in the Regge limit. 
  As we have not used the DGLAP evolution equations, which
assume the transverse momenta to be strictly ordered in
virtuality, we include important
logarithmic contributions which extend the realm of applicability.

\looseness=-1

\end{document}